\documentclass[prb,twocolumn,superscriptaddress,showpacs]{revtex4-1}

\pdfoutput=1

\usepackage{graphicx}% Include figure files
\usepackage{amsmath,amssymb}

\usepackage{color}
\usepackage{amssymb}
\usepackage{braket}

\def\diff{\mathrm d}
\def\mathi{\mathrm i}

\newcommand{\bx}{\ensuremath{\boldsymbol{\rho}}}
\newcommand{\by}{\ensuremath{\boldsymbol{G}}}
\newcommand{\bS}{\ensuremath{\boldsymbol{S}}}
\newcommand{\bU}{\ensuremath{\boldsymbol{U}}}
\newcommand{\bV}{\ensuremath{\boldsymbol{V}}}

\newcommand{\bK}{\ensuremath{\boldsymbol{K}}}

\newcommand{\bu}{\ensuremath{\boldsymbol{u}}}
\newcommand{\bv}{\ensuremath{\boldsymbol{v}}}

\newcommand{\wmax}{\ensuremath{\omega_\mathrm{max}}}

\newcommand{\tx}{\ensuremath{\tilde{x}}}
\newcommand{\ty}{\ensuremath{\tilde{y}}}
\newcommand{\tu}{\ensuremath{\tilde{u}}}
\newcommand{\tv}{\ensuremath{\tilde{v}}}
\newcommand{\tK}{\ensuremath{\tilde{K}}}

\begin{document}
\title{
Compressing Green's function using intermediate representation between imaginary-time and real-frequency domains}
\author{Hiroshi Shinaoka}
\affiliation{Department of Physics, Saitama University, 338-8570, Japan}
\author{Junya Otsuki}
\affiliation{Department of Physics, Tohoku University, Sendai 980-8578, Japan}
\author{Masayuki Ohzeki}
\affiliation{Graduate School of Information Sciences, Tohoku University, Sendai 980-8579, Japan}
\author{Kazuyoshi Yoshimi}
\affiliation{Institute for Solid State Physics, University of Tokyo, Chiba 277-8581, Japan}

\date{\today}

\begin{abstract}
New model-independent compact representations of imaginary-time data are presented in terms of the intermediate representation (IR) of analytical continuation.
%This is motivated by a recent numerical finding by the authors [J. Otsuki \textit{et al.}, arXiv:1702.03056].
We demonstrate the efficiency of the IR through continuous-time quantum Monte Carlo calculations of an Anderson impurity model.
We find that the IR yields a significantly compact form of various types of correlation functions.
This allows the direct quantum Monte Carlo measurement of Green's functions in a compressed form,
which considerably reduces the computational cost and memory usage.
Furthermore, the present framework will provide general ways to boost the power of cutting-edge diagrammatic/quantum Monte Carlo treatments of many-body systems.
\end{abstract}

\pacs{02.70.Ss}

\maketitle
\section{Introduction}
Many-body theories based on Matsubara Green's function are powerful tools to study correlated systems.
Elaborate diagrammatic methods have been widely used for investigating static and dynamic responses of the systems~\cite{Baym:1961do,Baym:1962dp,Bickers:1989hk}.
Modern quantum Monte Carlo (QMC) methods even provide access to numerically exact ground-state and dynamical properties of lattice models and impurity models~\cite{Werner:2006ko,Gull:2008cma,Gull:2011jda,Sandvik:1991ht,Rubtsov:2005iwb,Rombouts:1999tz,ProkofEv:1998gza,Beard:1996uw,Sandvik:1991ht,Iazzi:2015hi,2015PhRvB..91w5151W,Blankenbecler:1981gv,Kawashima:2004cla,Prokofev:1998fj}.
In these numerical calculations, however, one frequently faces two problems: (1) storage size and postprocessing cost of imaginary-time objects and (2) analytical continuation to the real-frequency axis.

The first issue becomes problematic in solving low-energy lattice models.
For instance, one needs to treat two-particle quantities for computing lattice susceptibilities.
Two-particle quantities also play a central role in some diagrammatic extensions of dynamical mean-field theory (DMFT)~\cite{Georges:1996un} for describing non-local spatial correlations~\cite{Toschi:2007fq,Rubtsov:2008cs}.
A recent technical advance is the compact representation of the imaginary-time dependence in terms of Legendre polynomials~\cite{Boehnke:2011dd}.
Efforts have been also made to describe the high-frequency asymptotic behavior of two-particle objects~\cite{Li:2016bn,2016arXiv161006520W,Kunes:2011is}.
However, the application of these elaborate methods to realistic models is still too computationally expensive.
A similar problem appears in quantum chemistry calculations based on a single-particle-level perturbative approach~\cite{Kananenka:2016cfa,Rusakov:2016eu}.
In this case, one needs to treat a much wider energy range than the low-energy models.
Thus, there is a high demand for a more compact representation as a key ingredient in cutting-edge simulations of many-body systems.
\begin{figure}
	\centering
	\includegraphics[width=0.3\textwidth,clip]{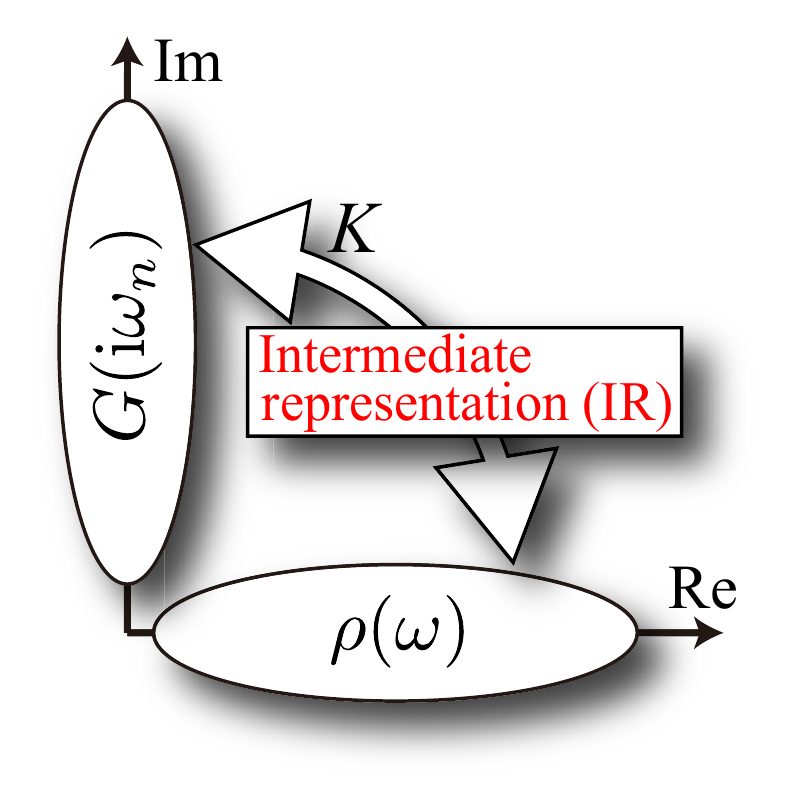}
	\caption{
		(Color online) Analytical continuation between real-frequency data $\rho(\omega)$ and Matsubara-frequency data $G(\mathi \omega_n)$ through the kernel $K$. The intermediate representation is defined in terms of a SVD of $K$.
	}
	\label{fig:IR}
\end{figure}

The second problem is \textit{ill-conditioned} 
analytical continuation from imaginary-time data to real-frequency axis.
One example is estimating the spectral function from imaginary-time Green's function computed in QMC simulations.
The problem can be formulated as the linear equation
\begin{align}
\by &= - \bK \bx,\label{eq:kernel}
\end{align}
where $\by$ and $\bx$ are vectors representing imaginary-time and real-frequency data and the matrix $\bK$ a kernel.
Since $\bK$ is ill-conditioned, the singular values of $\bK$ decay very fast.
As a result, most of independent components in $\bx$ give almost no contribution to $\by$.
Thus, if one simply minimizes $|\by + \bK \bx|^2$ with respect to $\bx$, 
any errors in $\by$ are enormously amplified in $\bx$.

The authors have recently developed a new method for analytical continuation of QMC data~\cite{Otsuki2017}.
We demonstrated that, using a modern information theory called ``sparse modeling", relevant information can be successfully extracted from imaginary-time data with statistical errors.
One of the key steps in this method is to transform the original data into a basis obtained by the singular value decomposition (SVD) of the matrix $\bK$.
As a result, after the errors are properly removed, the original data are expressed with only a few components.
A similar observation was made in previous studies where SVD was employed in the context of analytical continuation~\cite{CREFFIELD:1995vh,Jarrell:1996is,Gunnarsson:2010jt}.
These strongly suggest a possibility that this basis, which plays a key role in the analytical continuation, settles the first issue on the storage size and computational cost.

In this paper, we show that this model-independent basis can be used to compress various types of imaginary-time objects.
We coin the term ``intermediate representation (IR)" for this basis as it is defined between real-frequency and imaginary-time domains~(Fig.~\ref{fig:IR}).
After investigating the properties of the IR in detail,
we assess its efficiency for a single-site Anderson impurity model through continuous-time QMC simulations.
We thus demonstrate that the IR provides significantly compact representations of
the single-particle Green's function, the charge susceptibility and the generalized susceptibility.

\section{Properties of basis functions}
We start our discussion by considering the spectral (Lehmann) representation of a single-particle Green's function $G$
\begin{align}
G(\tau) &= -\int_{-\wmax}^{\wmax} d\omega K(\tau, \omega) \rho(\omega),\label{eq:fwd}
\end{align}
where we take $\hbar = 1$ and $0 \le \tau  \le \beta$.
This equation is reduced to Eq.~(\ref{eq:kernel}) when the variables $\tau$ and $\omega$ are discretized.
The spectra function $\rho(\omega)$ is given by 
\begin{align}
\rho_\mathrm{F}(\omega) &= -\frac{1}{\pi}\mathrm{Im} G(\omega + \mathi 0),\label{eq:rhoF}
\end{align}
or
\begin{align}
\rho_\mathrm{B}(\omega) &= -\frac{1}{\pi\omega}\mathrm{Im} G(\omega + \mathi 0),\label{eq:rhoB}
\end{align}
in the fermionic/bosonic case, respectively.
The kernel is defined correspondingly 
\begin{align}
   K_\mathrm{F}(\tau, \omega) &\equiv \frac{e^{-\tau\omega}}{1 + e^{-\beta \omega}},\label{eq:KF}
\end{align}
or
\begin{align}
K_\mathrm{B}(\tau, \omega) &\equiv \omega \frac{e^{-\tau\omega}}{1 - e^{-\beta \omega}}.\label{eq:KB}
\end{align}
The extra $\omega$'s in Eqs.~(\ref{eq:rhoB}) and (\ref{eq:KB}) are introduced to avoid a singularity of the kernel at $\omega=0$.
We assume that the spectral function is bounded in the interval $[-\wmax,~\wmax]$.
Note that there is a similar spectral representation for the self-energy of a system of fermions~\cite{Luttinger:1961wb}.

\begin{figure}
	\centering
	\includegraphics[width=0.5\textwidth,clip]{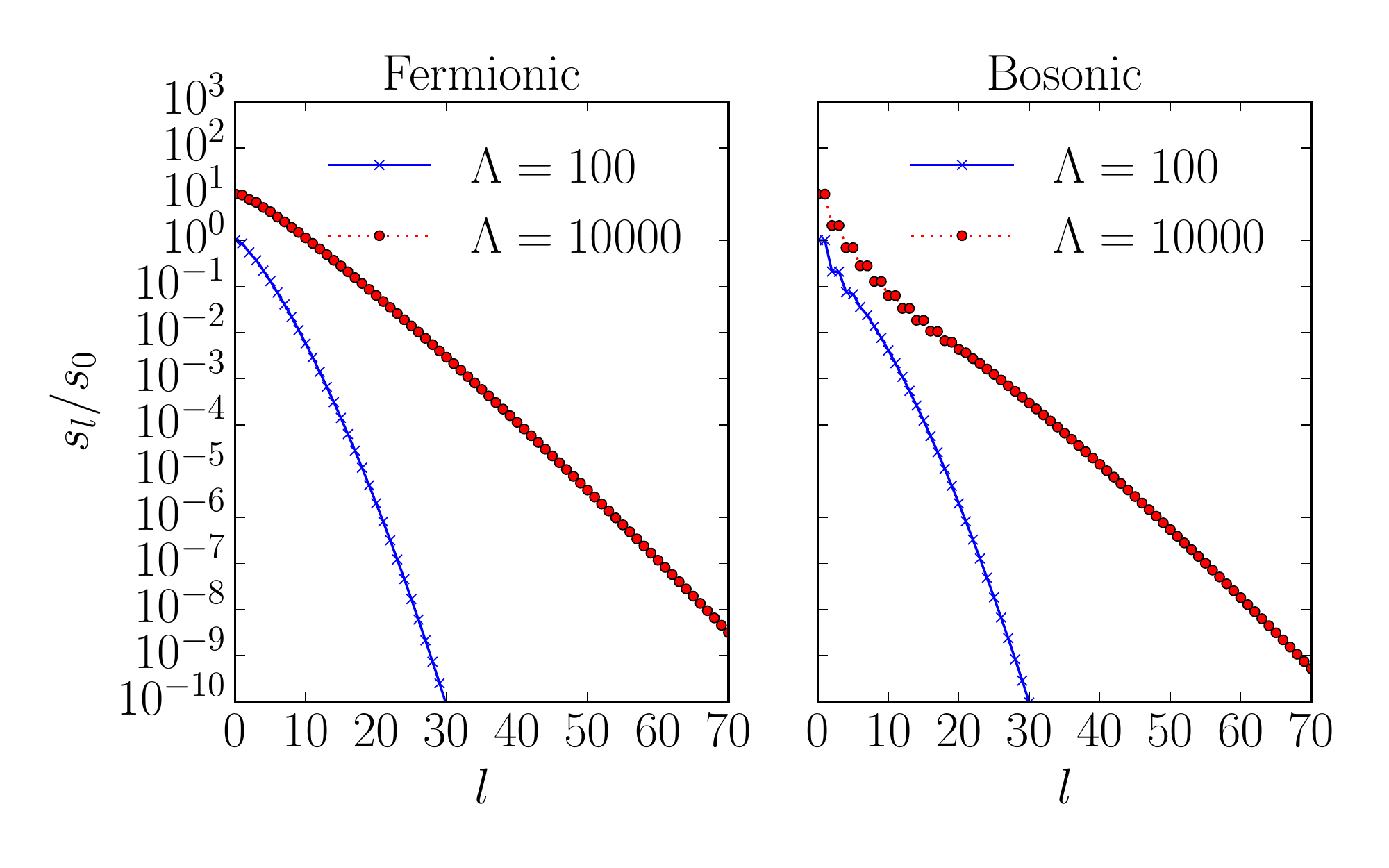}
	\includegraphics[width=.5\textwidth,clip]{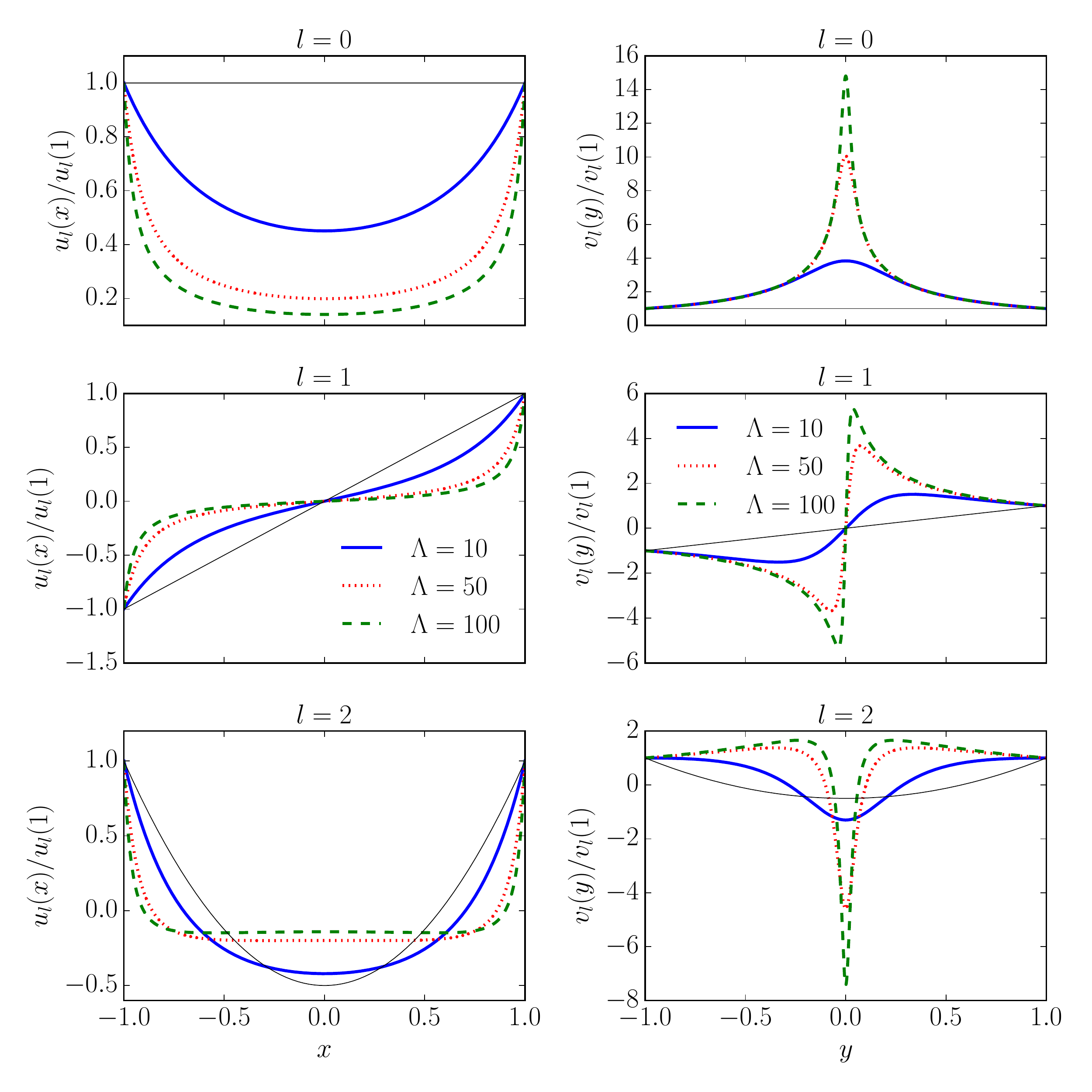}
	\caption{
		(Color online) Upper panel shows the singular values computed for the fermionic and bosonic kernels.
		The data for $\Lambda=10000$ are multiplied by a constant for better readability.
		The lower panel shows the basis functions in the imaginary-time domain [$u_l(x)$] and in the real-frequency domain [$v_l(y)$] computed for the fermionic kernel. 
		The solid gray lines show Legendre polynomials.
	}
	\label{fig:svd}
\end{figure}

For convenience, we transform the variables $\tau$ and $\omega$ into dimensionless variables $x \equiv 2\tau/\beta -1 \in [-1, 1]$ and $y\equiv \omega/\wmax \in [-1, 1]$.
The kernels are then rewritten as
\begin{align}
K_\mathrm{F}(x, y)&=\frac{e^{-\frac{\Lambda}{2} x y}}{\cosh(\frac{\Lambda}{2} y)},\\
K_\mathrm{B}(x, y)&=y \frac{e^{-\frac{\Lambda}{2} x y}}{\sinh(\frac{\Lambda}{2} y)}.\label{eq:kB}
\end{align}
Here we introduced a dimensionless parameter $\Lambda\equiv \beta \wmax$.
The IR changes its form depending on the value of $\Lambda$ as we will see below. 

The IR is now defined through the decomposition of the kernels as
\begin{align}
  K(x, y) &= \sum_{l=0}^{\infty} s_l u_l(x) v_l(y).\label{eq:kernel-exp}
\end{align}
This decomposition can be performed by SVD of a kernel matrix $\bK$ defined on a dense uniform mesh:
$\bK = \bU \bS \bV^\dagger$.
In the continuous limit, column vectors of $\bU$ ($\bV$) yield orthonormal basis set $\{u_l(x)\}$ in the $\tau$ domain ($\{v_l(y)\}$ in the $\omega$ domain)
~\footnote{
	For a large $\Lambda$, some of zeros of $\{u_l(x)\}$ and $\{v_l(y)\}$ are distributed close to $x=\pm 1$ and $y=0$, respectively. For uniform meshes, this leads to a slow convergence of results with respect to the number of mesh points. To improve the convergence, we actually use a non-uniform mesh. Note that special care must be taken in the transformation of $\bK$ to make sure that $\bU$ and $\bV$ converge to the same $\{u_l(x)\}$ and $\{v_l(y)\}$ obtained by uniform meshes. A technical note is given in the Appendix.}.
$s_l$ ($>0$) are singular values in non-ascending order.
In the literature, SVD was utilized in the context of analytical continuation~\cite{CREFFIELD:1995vh,Jarrell:1996is,Gunnarsson:2010jt}.
Our idea is to \textit{represent} imaginary-time dependence by $u_l(x)$ to acquire compact forms of correlation functions.

We expand a given imaginary-time object $G(\tau)$ and the corresponding spectral function $\rho(\omega)$, respectively, in terms of $\{ u_l(x) \}$ and $\{ v_l(y) \}$ as
\begin{align}
  G(\tau) &= \frac{\sqrt{2}}{\beta}\sum_{l\geq 0} g_l u_l(\beta(x+1)/2),\label{eq:IR-exp}\\
\rho(\omega) &= \sum_{l\geq 0} \rho_l v_l(\wmax y).\label{eq:IR-exp2}
\end{align}
Using Eqs.~(\ref{eq:fwd}) and (\ref{eq:kernel-exp}), we can demonstrate that the coefficients $g_l$ and $\rho_l$ have one-to-one correspondence
\begin{align}
g_l = -s_l \rho_l.
\end{align}
Note that the singular values $\{s_l\}$ decay at least exponentially as shown in the upper panel of Fig.~\ref{fig:svd}.
It leads to an exponential decay of $g_l$, provided that $\rho_l$ does not grow for large $l$.
In practical cases, we have confirmed that $\rho_l$ vanishes as $l$ increases, and $g_l$ decays even faster than $s_l$. 
The expansion in Eq.~(\ref{eq:IR-exp}), therefore, may be truncated at a certain order, which will be demonstrated later using QMC simulations. 

Here, we investigate the properties of the IR basis to get intuitive understanding why the expansion converges fast. 
Figure~\ref{fig:svd} shows the basis functions computed for the fermionic case.
The real-frequency basis functions $v_l(y)$ have fine structure around $\omega=0$, which becomes shaper as $\Lambda$ is increased.
This is consistent with that the kernel does not filter out the fine structure of a spectral function at small $\omega$.
For $u_l(x)$, two notable features are clearly discernible: $u_l(x)$ is an even/odd function for even/odd $l$, and there are $l$ zeros~\footnote{A previous report on the number of zeros can be found in Ref.~\onlinecite{CREFFIELD:1995vh}.}.
More importantly, we found that $u_l(x)$ [and $v_l(y)$] converges to the $l$-th Legendre polynomial $P_l(x)$ up to a normalization factor as $\Lambda\rightarrow 0$~\footnote{A note is given in the Appendix. In the note, we numerically and algebraically show that the first few basis functions converge to the Legendre polynomials in the high temperature limit.}.
It means that our representation using the IR basis includes the Legendre representation as a special limit.
However, since this limit corresponds to the high-$T$ limit, the Legendre expansion may not be efficient for low $T$.
A difference between the IR basis and the Legendre polynomials becomes clear as $\Lambda$ is increased:
The values of $u_0(x)$ and $u_1(x)$ change more rapidly around $x=\pm 1$, which resemble the behavior of diagonal and off-diagonal elements of the Green's function, respectively.
Moreover $u_0(0)$ becomes suppressed, similarly to the low-$T$ behavior of the diagonal elements $G(\tau=\beta/2) \propto T$.
Therefore, an efficient descriptions in terms of the IR basis is expected especially at low $T$. 

\section{Results of quantum Monte Carlo simulations}
\subsection{Model}
Now we demonstrate the efficiency of the IR for describing various types of imaginary-time objects.
As a simple example, we consider the particle-hole symmetric single-site Anderson impurity model defined by the Hamiltonian
\begin{eqnarray}
\mathcal{H} &=& - \mu  \sum_\sigma c^\dagger_{\sigma} c_{\sigma} + U n_{\uparrow} n_{\downarrow} + \sum_{k\sigma} (c^\dagger_{\sigma} a_{k \sigma} +a^\dagger_{k \sigma} c_{\sigma} )\nonumber\\
&& + \sum_{\alpha} \sum_{k \sigma} \epsilon_k a^\dagger_{k \sigma} a_{k \sigma}\label{eq:model}
\end{eqnarray}
with $\mu=U/2$ and $\sigma$ is spin index.
$c_\sigma$ and $c^\dagger_\sigma$ are annihilation and creation operators at the impurity site, while $a_{k\sigma}$ and $a^\dagger_{k\sigma}$ are those of the bath sites ($k$ is the internal degree of freedom of the bath).
The distribution of $\epsilon_k$ is a semicircular density of states of width $4$.
We solve the model and compute correlation functions by means of the hybridization expansion continuous-time Monte Carlo technique~\cite{Werner:2006ko}.

\subsection{Single-particle Green's function}
First, we discuss the impurity single-particle Green's function defined as $G_\sigma(\tau) = -\braket{c_\sigma(\tau) c^\dagger_\sigma(0)}$ ($0 \le \tau \le \beta$). 
We expand $G_\sigma(\tau)$ in terms of an orthogonal basis set $\{f_l(x)\}$ [$P_l$ or $u_l$] as
\begin{align}
   G_\sigma(\tau) &=\frac{\sqrt{2}}{\beta}  \sum_{l\ge 0} G^\sigma_l\frac{ f_l(x(\tau))}{\sqrt{N_l}},\label{eq:exp}
\end{align}
where $x(\tau) = 2\tau/\beta - 1$ and $\int_{-1}^{1} f_l(x) f_{l^\prime}(x)dx = N_l \delta_{ll^\prime}$.
We directly measure the coefficients $G_l^\sigma$ in QMC simulations as described in Ref.~\onlinecite{Boehnke:2011dd}.

\begin{figure}
	\centering
	\includegraphics[width=0.5\textwidth,clip]{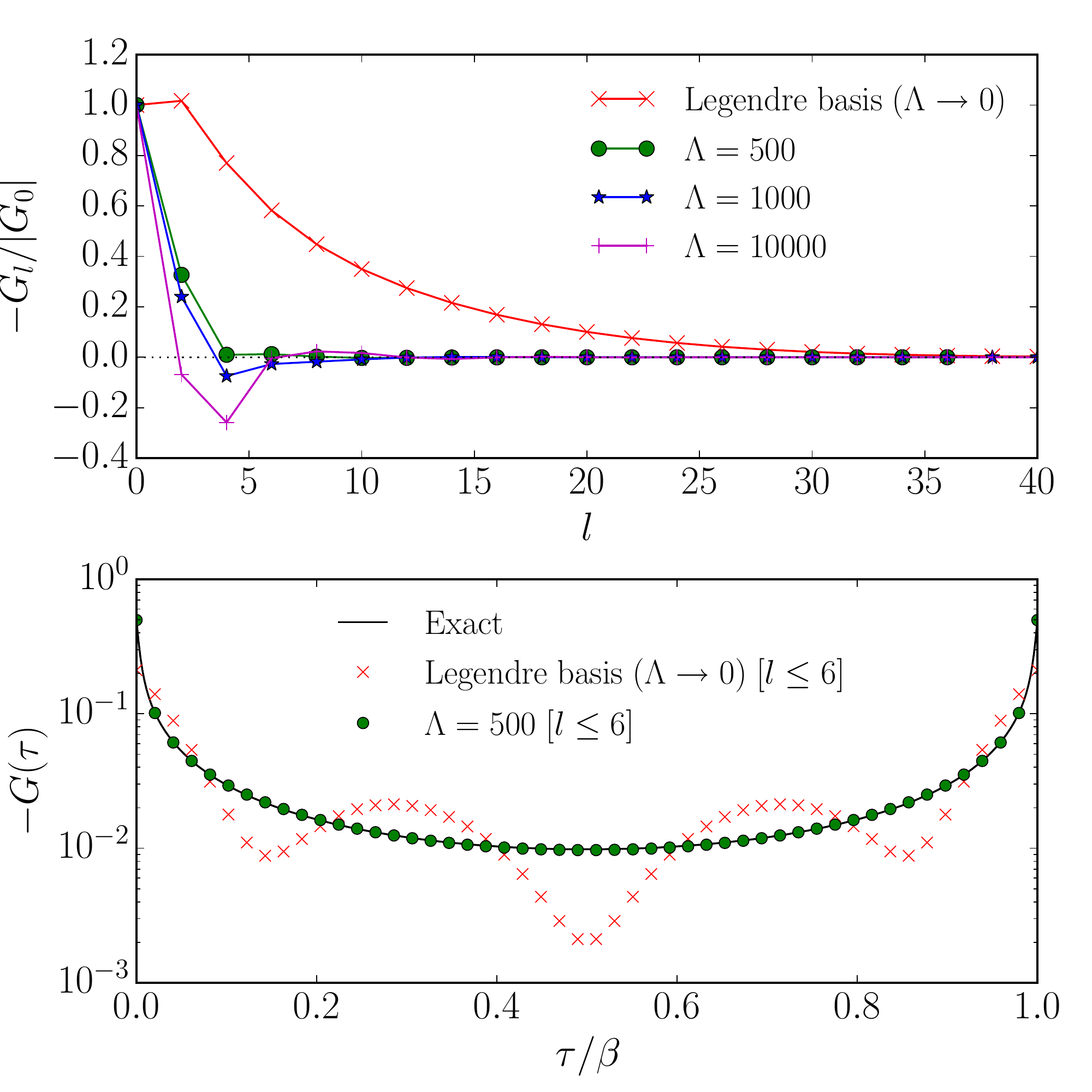}
	\caption{
		(Color online)
		Single-particle Green's function computed for the model (\ref{eq:model}) with $U=4$ and $\beta=100$.
		Upper panel: Expansion coefficients $G_l$.
		We show only data for even $l$ since $G_l$ for odd $l$ are zero due to the particle-hole symmetry.
		Lower panel: $G(\tau)$ reconstructed from a few small-$l$ coefficients.
		All the data are averaged along the spin index. 
	}
	\label{fig:G1}
\end{figure}

In Fig.~\ref{fig:G1}, we show the coefficients $G_l$ obtained for $U=4$ and $\beta=100$.
The large-$l$ asymptomatic behavior of the Legendre representation is known to be exponential~\cite{Boehnke:2011dd}, while the Matsubara-frequency representation has a $1/\mathi \omega_n$ tail.
As expected,
the IR yields coefficients decaying even faster than the Legendre basis.
One can expect that the most compact representation is obtained when $\Lambda/\beta$ matches the actual width of the spectrum.
This suggests a practical way to choose an appropriate value of $\Lambda$.
Actually, the optimal value obtained is $\Lambda\simeq 1000$ for $\beta=100$, being consistent with the largest dimensionless energy scale of the system, i.e., $\beta U$, $\beta W = 400$.
As $\Lambda$ exceeds the optimal value, the efficiency gets worse only slowly.
In particular, we observed the non-monotonic behavior of $G_l$ around $l=5$ for $\Lambda > 500$,
which signals that $\Lambda$ exceeds an optimal value.

In Fig.~\ref{fig:G1}, we also show $G(\tau)$ reconstructed from the coefficients for $l\le 6$.
The data obtained by the IR ($\Lambda=500$) shows a perfect agreement with the numerically exact data, while the truncation in the Legendre representation results in large Gibbs oscillations.

\subsection{Charge susceptibility}
\begin{figure}
	\centering
	\includegraphics[width=0.5\textwidth,clip]{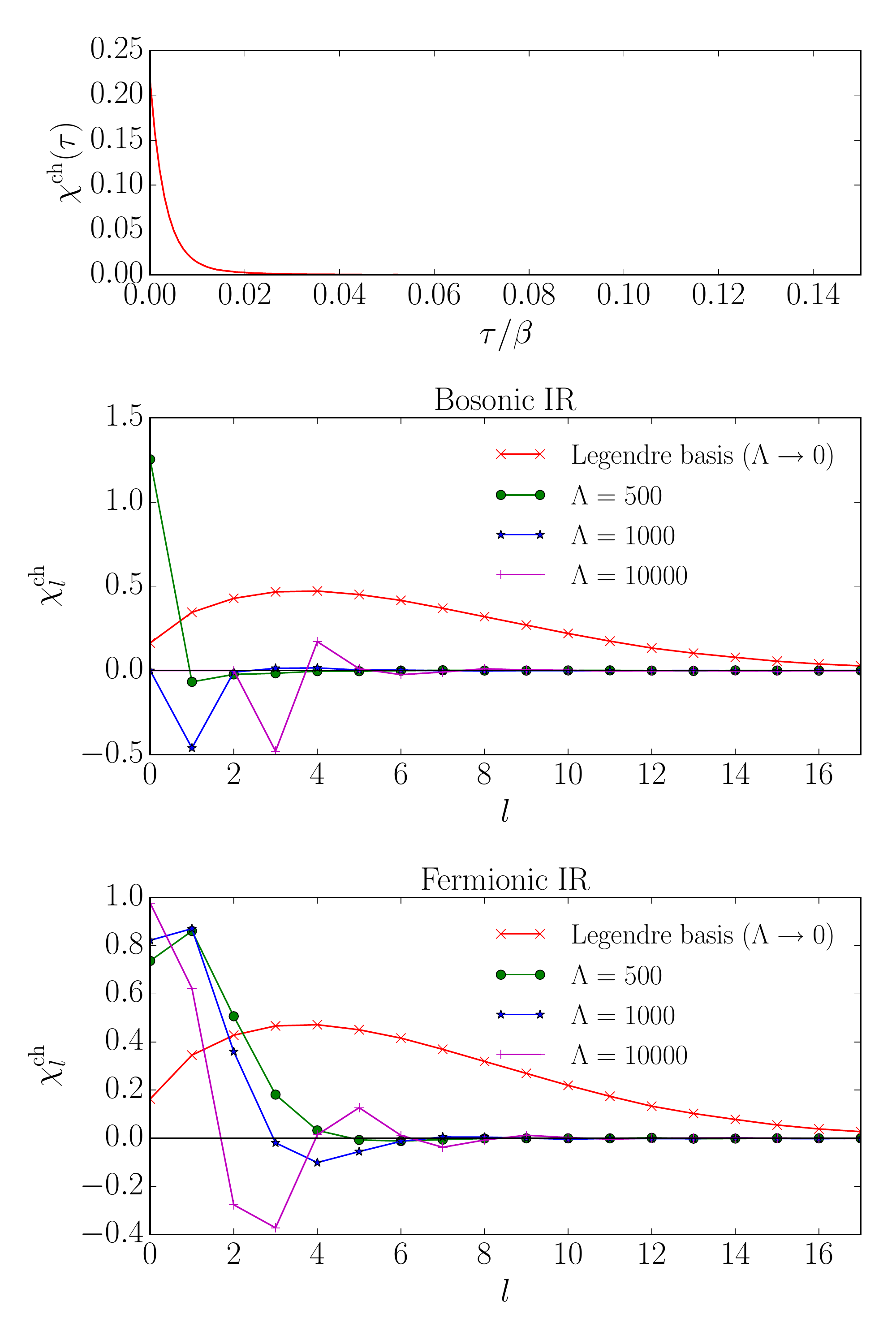}
	\caption{
		(Color online)
		Charge susceptibility computed for the model (\ref{eq:model}) with $U=4$ and $\beta=100$.
		The upper panel shows the $\tau$ dependence.
		The middle and lower panels show the coefficients in terms of the bosonic and fermionic IR's.
	}
	\label{fig:chi}
\end{figure}

Second, as a typical bosonic quantity, we analyze the charge susceptibility $\chi^\mathrm{ch}(\tau)$ defined by
\begin{align}
 \chi^\mathrm{ch}(\tau) &= \braket{n(\tau) n(0)} - \braket{n}^2,
\end{align}
on the interval $[0,\beta]$ ($n_\sigma \equiv \sum_{\sigma} c^\dagger_\sigma c_\sigma$).
We expand the $\tau$ dependence using Eq.~(\ref{eq:exp}) in terms of the bosonic IR or the fermionic IR.
Strange as the latter may sound, it is possible since the basis functions $v_l(x)$ always form a complete basis set on the interval $[-1, 1]$.
Figure~\ref{fig:chi} shows the results obtained for $\beta=100$.
Remarkably, the bosonic IR requires only few coefficients beyond statistical errors.
Again, the most compact representation is obtained when $\Lambda/\beta$ matches the spectral width.
On the other hand, surprisingly, the fermionic IR is better than the Legendre basis [the lower panel of Fig.~\ref{fig:chi}].
However, the most compact representation is obtained with the bosonic IR with $\Lambda\simeq 500$.
This shows the importance of using the bosonic IR for bosonic quantities.
 % since the bosonic IR is tailor-made for bosonic quantities.

\subsection{Two-particle Green's function}
Finally, we demonstrate that the IR's for single-particle Green's functions can be used for expanding objects with multiple time indices.
As an example, we consider the generalized susceptibility defined by
\begin{align}
 & \tilde{\chi}(\tau_{14}, \tau_{24}, \tau_{34})\equiv \langle T_\tau c^\dagger_{\sigma_1}(\tau_1) c_{\sigma_2}(\tau_2)  c^\dagger_{\sigma_3} (\tau_3) c_{\sigma_4} (\tau_4)\rangle \nonumber\\
 & \hspace{2em} -  \langle T_\tau c^\dagger_{\sigma_1}(\tau_1) c_{\sigma_2}(\tau_2) \rangle \langle T_\tau c^\dagger_{\sigma_3} (\tau_3) c_{\sigma_4} (\tau_4)\rangle
 ,\label{eq:G2}
\end{align}
where $\tau_{ab} \equiv \tau_a - \tau_b$ and the second term subtracts the trivial contribution of the bubble diagram.
In Ref.~\onlinecite{Boehnke:2011dd}, Boehnke \textit{et al.} introduced the mixed representation of Legendre polynomials and bosonic Matsubara frequencies, in which $\tau_{12}$ and $\tau_{34}$ dependence of a two-particle object is expanded in terms of the Legendre polynomials, while the $\tau_{14}$ dependence is described through Fourier modes $e^{\mathi \omega_m \tau_{14}}$~\cite{Boehnke:2011dd}.
This is motivated by that fact that the Bethe-Salpeter equation is diagonal in the bosonic frequency $\mathi \omega_m$ that is connected to $\tau_{14}$ through a Fourier transformation.

We now simply replace the Legendre polynomials by the IR for the fermionic kernel. % in the mixed representation.
This leads to
\begin{align}
%& \tilde{\chi}_{\sigma_1 \sigma_2 \sigma_3 \sigma_4}(\tau_{12}, \tau_{34}, \tau_{14}) \equiv \sum_{ll^\prime \ge 0} \sum_{m \in \mathcal{Z}} 2\beta^{-3} (-1)^{l^\prime+1} \nonumber\\
& \tilde{\chi}_{\sigma_1 \sigma_2 \sigma_3 \sigma_4}(\tau_{12}, \tau_{34}, \tau_{14}) \equiv \sum_{ll^\prime \ge 0} \sum_{m \in \mathcal{Z}} \sqrt{\frac{2}{N_l}}\sqrt{\frac{2}{N_{l^\prime}}} \beta^{-3} (-1)^{l^\prime+1} \nonumber\\
& \hspace{2em} u_l[x(\tau_{12})] u_{l^\prime}[x(\tau_{34})] e^{\mathi \omega_m \tau_{14}} \tilde{\chi}^{\sigma_1 \sigma_2 \sigma_3 \sigma_4}_{ll^\prime}(\mathi \omega_m).\label{eq:G2-mixed}
\end{align}
Hereafter, we consider only the spin diagonal components $\tilde{\chi}_{\uparrow \uparrow \uparrow \uparrow}~(=\tilde{\chi}_{\downarrow \downarrow \downarrow \downarrow})$ and drop the spin indices.
In practice, we measure the coefficients of the first term in Eq.~(\ref{eq:G2}) in QMC simulations and subsequently subtract the bubble-diagram contribution (second term) computed from the data of the single-particle Green's function.
The accumulation of the coefficients for a single bosonic frequency requires $O(k^4 N^2)$ operations,
where $k$ is the expansion order of QMC and $N$ is the number of the IR basis functions or Legendre polynomials for fermionic frequencies.
Thus, any reduction of $N$ will significantly speed-up QMC measurement.
We refer the readers to Ref.~\onlinecite{Boehnke:2011dd} for more technical details.
 %$\tilde{\chi}_{ll^\prime}(\mathi \omega_m)$ 
 %directly 

\begin{figure}
	\centering
	\includegraphics[width=0.5\textwidth,clip]{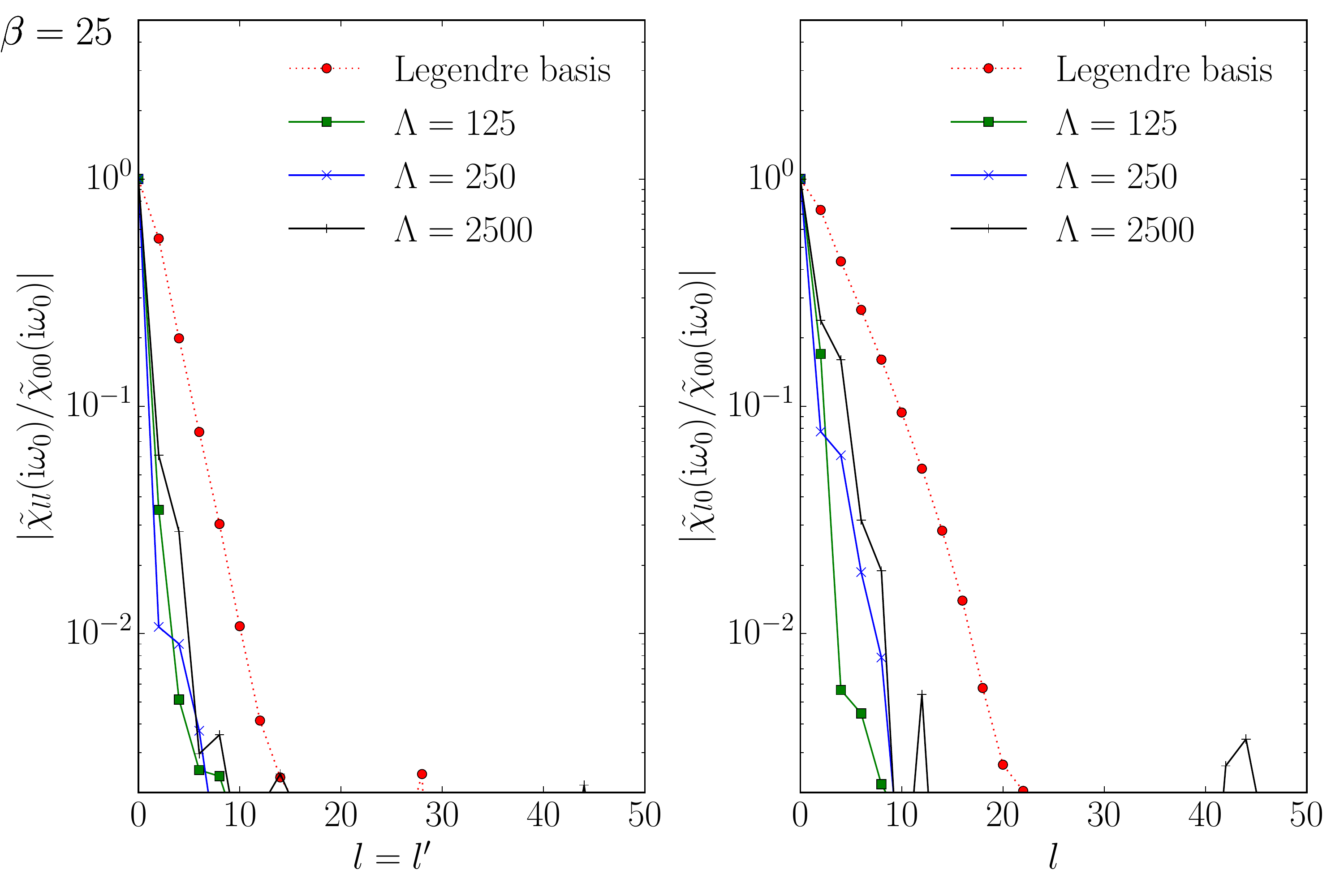}
	\includegraphics[width=0.5\textwidth,clip]{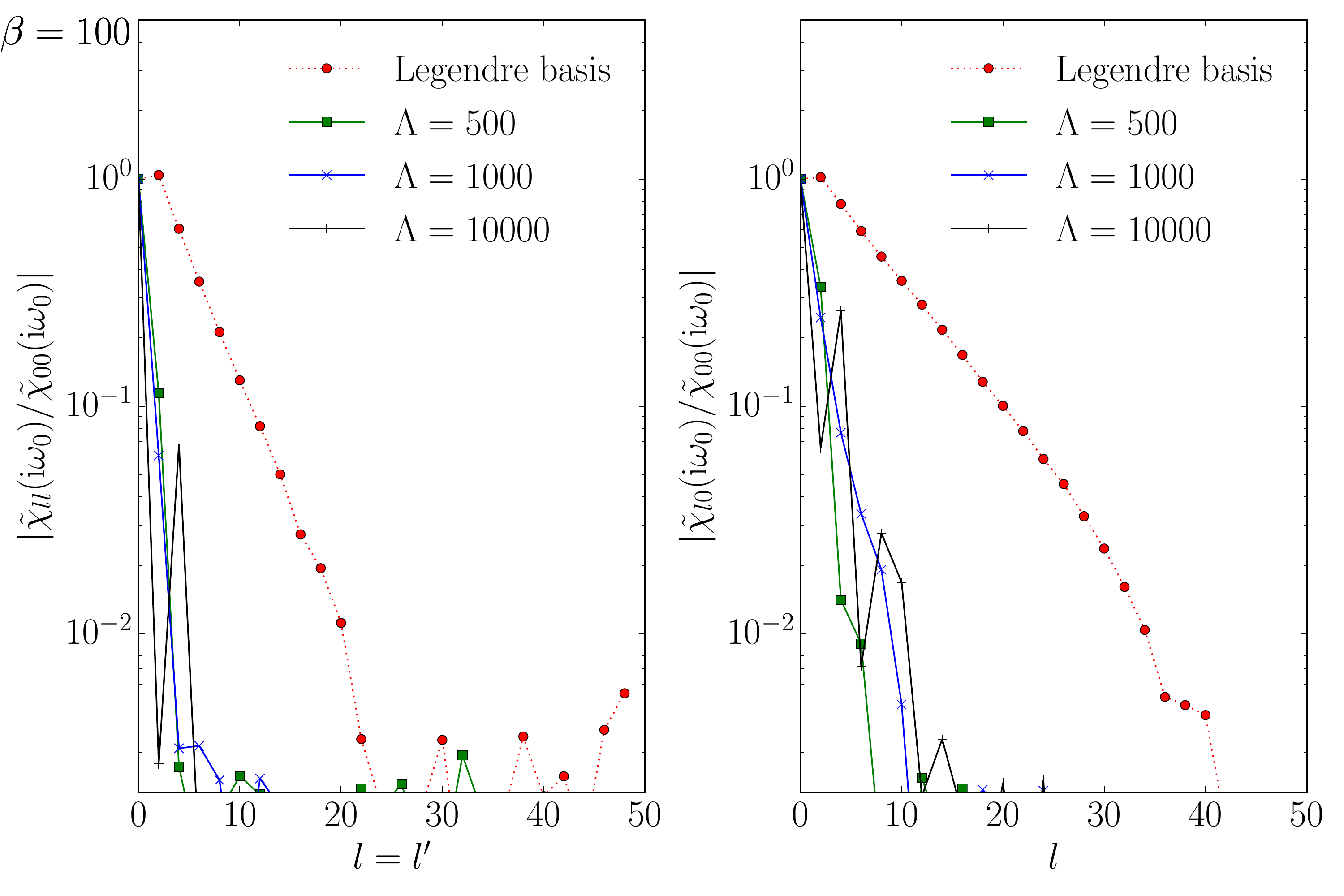}
	\caption{
		(Color online) Coefficients of generalized spin susceptibilities of the model (\ref{eq:model}) measured in the mixed representation of bosonic Matsubara frequencies and the fermionic IR.
		We plot data along the diagonal line $l=l^\prime$ (left panel) and those along the line $l^\prime=0$ (right panels).
		Upper and lower panels show data for $\beta=25$ and 100, respectively.
%		The upper panel shows $|\chi_{ll^\prime}/\chi_{00}|$ at the boson frequency $\omega_m = 0$.
%		In the lower panels, we plot the data for $l=l^\prime$ (left lower panel) and $l^\prime=0$ (right lower panel).
		We subtract the contributions of the bubble diagram from the data and show data for even $l$ and $l^\prime$. 
	}
	\label{fig:G2}
\end{figure}

We show the results at the zero bosonic frequency $\mathi \omega_m=0$ for $\beta=25$ and $100$ in Fig.~\ref{fig:G2}.
In the mixed representation of the IR, the coefficients for even $l$ and $l^\prime$ show a much faster decay than those for the Legendre basis.
The other coefficients for odd $l$ or odd $l^\prime$ take on a very small value, compatible with a vanishing value within statistical errors.
As $\beta$ increases, the decay becomes slower for the Legendre basis.
On the other hand, for the IR basis,
the rate of the decay depends on the value of $\Lambda$ slightly.
But one does not observe a noticeable slowdown in the decay if the value of $\Lambda$ is chosen appropriately.
As a result, the new basis becomes more superior as $\beta$ increases.
%In the lower panel of Fig.~\ref{fig:G2}, we show the data on the diagonal line $l=l^\prime$ and those on the line $l^\prime=0$.

An interesting observation is that the most compact representation is obtained for a value of $\Lambda$ close to the optimal one for the single-particle Green's function for $\beta=100$.

\begin{figure}
	\centering
	\includegraphics[width=0.5\textwidth,clip]{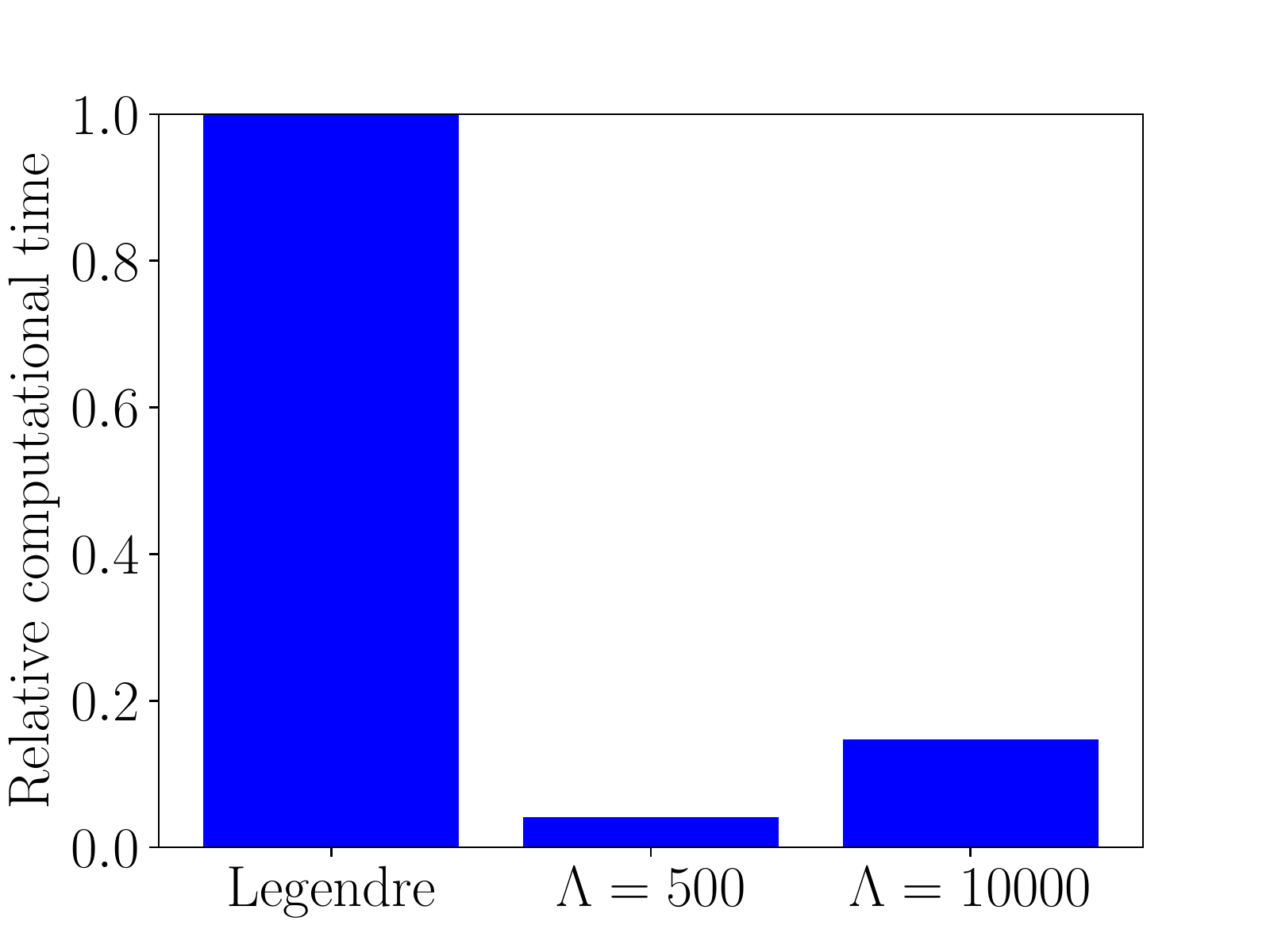}
	\caption{
		(Color online) Relative computational time of measurement of two-particle Green's function at bosonic frequency $\omega_m=0$ for $\beta=100$.
		We employed 50 (Legendre), 10 ($\Lambda=500$), 20 ($\Lambda=10000$) basis functions in the measurement.
	}
	\label{fig:G2-timing}
\end{figure}

To demonstrate advantages of the new basis functions in practical QMC calculations,
we measured the computational time of the measurements of two-particle Green's function.
The results for $\beta=100$ are shown in Fig.~\ref{fig:G2-timing}.
We employed 50 (Legendre), 10 ($\Lambda=500$), 20 ($\Lambda=10000$) basis functions in the measurement so that the data beyond the noise level are accumulated (see Fig.~\ref{fig:G2}).
The simulations were performed on a 2.5GHz Intel Xeon CPU (E5-2680 v3) without parallelization.
One can see that the measurement for $\Lambda=500$ is faster than the case of the Legendre polynomials approximately by 24 times,
being consistent with the expected scaling.

\section{Summary}
In summary, we proposed the new compact representations of imaginary-time data, which was named IR, through the lenses of analytical continuation.
The new basis does not depend on the details of the systems.
In particular, we studied the properties of the IR's for fermionic and bosonic Green's functions.
We found that the conventional Legendre basis corresponds to the high-$T$ limit of the IR.
The IR was applied to QMC simulations of the single-site quantum impurity model.
We confirmed that the present method yields significantly compact form of various imaginary-time correlation functions than the conventional ones.
This allows the direct measurement of Green's functions in a compressed form,
which reduces the computational cost and memory usage.

An optimal value of $\Lambda$ depends on temperature and the width of spectral function.
The numerical tests indicate that the data remain compact even when $\Lambda$ exceeds the optimal value.
Thus, one does not have to tune the value of $\Lambda$ very precisely.
It may be practically efficient enough to let the value of $\Lambda$ be on the large side.

The present scheme provides a new approach to solve technical issues in a variety of state-of-the-art treatments of many-body quantum systems.
For instance, one may be able to perform diagrammatic calculations with Bethe-Salpeter/parquet equations in the IR.
Promising applications are the diagrammatic extensions of DMFT (dual fermions~\cite{Toschi:2007fq} and dynamical vertex approximation~\cite{Rubtsov:2008cs}) and the computation of lattice susceptibilities within DMFT.
On the other hand, the kernel for the Keldysh Green's function is also known to be ill-conditioned~\cite{Dirks:2013fr}.
The present scheme may be easily applied to non-equilibrium cases.
Furthermore, a modern regularization technique will enable to separate relevant information from statistical noise in the IR~\cite{Otsuki2017}.
\begin{acknowledgments}
	We are grateful to Emanuel Gull, Kristjan Haule, Yusuke Nomura and Philipp Werner for their useful comments on the manuscript.
	HS thanks fruitful discussions with Lewin Boehnke on the Legendre basis..
	HS was supported by JSPS KAKENHI Grant No. 16H01064 (J-Physics), 16K17735.
	JO was supported by JSPS KAKENHI Grant No. 26800172, 16H01059 (J-Physics).
	MO was supported by MEXT KAKENHI Grant No. 25120008, JST CREST and JSPS KAKENHI No. 16H04382.
	KY was supported by Building of Consortia for the Development of Human Resources in Science and Technology, MEXT, Japan.
	Part of the calculations were run on the ISSP supercomputing system with codes based the ALPSCore libraries~\cite{Gaenko:2016ic} and ALPSCore/CT-HYB~\cite{2017CoPhC.215..128S}.
	
\end{acknowledgments}

\bibliography{ref,unpublish}

\appendix
\section{Decomposition of the kernel}
\subsection{Singular value decomposition by means of a linear mesh}
We start our discussion with
\begin{align}
K(x, y) &= \sum_{l=0}^{\infty} s_l u_l(x) v_l^*(y).\label{eq:kernel-exp-suppl}
\end{align}
This equation is recast into
\begin{align}
s_l &= \int \diff x \diff y~u_l^*(x) K(x, y) v_l(y).~\label{eq:int}
\end{align}
Now, we introduce equally-spaced $N$ points $x_n$ and $y_m$ on the interval $[-1, 1]$ ($n,m = 0, \cdots, N-1$).
We approximate $u_l(x)$ and $v_l(y)$ as
\begin{align}
	u_l(x) &= \frac{2}{N} \sum_{n=0}^{N-1} u_{l,n} \delta(x - x_n),\\
    v_l(y) &= \frac{2}{N} \sum_{m=0}^{N-1} v_{l,m} \delta(y - y_m).\\	
\end{align}
Substituting these into Eq.~(\ref{eq:int}) leads to
\begin{align}
s_l &= \boldsymbol{u}_l^\dagger \boldsymbol{K} \boldsymbol{v}_l,
\end{align}
where $\boldsymbol{u}_l$ and $\boldsymbol{v}_l$ are the column vectors whose elements are $u_{l,n}$ and $v_{l,m}$, respectively.
For $N\gg 1$, the orthonormal condition of $u_l(x)$ and $v_l(y)$ is equivalent to that of the column vectors. 
The matrix element of $\boldsymbol{K}$ at $(n,m)$ is given by $\frac{4}{N^2}K(x_n, y_m)$.
Such vectors can be computed by a singular value decomposition (SVD) of the matrix $\bK$ as
\begin{align}
\bK &= \bU \bS \bV^\dagger,
\end{align}
where $\bS$ is the diagonal matrix whose elements are given by $s_l$, 
$\bU= (\bu_0, \cdots, \bu_{N-1})$ and $\bV= (\bv_0, \cdots, \bv_{N-1})$.

\subsection{Double-exponential mesh}
The linear mesh is not optimal because the values of the basis functions change very rapidly around $x=\pm 1$ and $y=0$.
Let us consider the change of variables $x = f(\tx)$ and $y = g(\ty)$.
Then, Eq.~(\ref{eq:kernel-exp-suppl}) reads
\begin{align}
s_l &= \int \diff x \diff y~u_l^*(x) K(x, y) v_l(y)\nonumber\\
&= \int \diff \tx \diff \ty~\tu_l^*(\tx) \tK(\tx, \ty) \tv_l(\ty),\label{eq:kernel-exp2}
\end{align}
where 
\begin{align}
  \tu_l(\tx) &\equiv \sqrt{f^\prime(\tx)}u_l(f(\tx)), \\
  \tv_l(\ty) &\equiv \sqrt{g^\prime(\ty)}u_l(g(\ty)), \\
  \tK(\tx, \ty) &\equiv \sqrt{f^\prime(\tx)g^\prime(\ty)} K(f(\tx), g(\ty)).
\end{align}
Here, we assume $f^\prime(\tx)\ge 0$ and $g^\prime(\ty) \ge 0$.
Note that $\tu_l(\tx)$ and $\tv_l(\ty)$ are orthonormal functions with respect to $\tx$ and $\ty$, respectively.
One can also solve Eq.~(\ref{eq:kernel-exp2}) in a way analogous to the solution of Eq.~(\ref{eq:kernel-exp2}).

This provides the possibility of choosing appropriate transformations so that we have more dense points in the regions where the values of the basis functions change rapidly, i.e., $x=\pm 1$ and $y=0$.
In particular, we adopt the double-exponential transformations~\cite{Takahasi:1974jt}
\begin{align}
   x &= \tanh\left(\frac{\pi}{2} \sinh \tx\right),\\
   y &= \frac{1}{2} \left\{\tanh\left(\frac{\pi}{2} \sinh \ty\right) + 1\right\}~(y > 0),\\
   y &= \frac{1}{2} \left\{\tanh\left(\frac{\pi}{2} \sinh \ty\right) - 1\right\}~(y < 0)
\end{align}
with the cutoff $|\tx| \le 4$ and $|\ty| \le 4$.
This transformation maps $x \in [-1, 1]$ to $\tx \in [-\infty, \infty]$.
The cutoff can be introduced very safely because the derivative of $f(\tx)$ and $g(\ty)$ show a double-exponential decay.
We found that $N=1001$ gives sufficiently accurate solutions for our purpose.
For more technical details, please study the Python scripts provided in the supplemental material.

\section{Asymptotic behavior of basis functions of Intermediate representation (IR)}
\begin{figure*}
	 \begin{tabular}{c}
		\begin{minipage}{0.5\hsize}
			\centering
			\includegraphics[width=0.95\textwidth,clip]{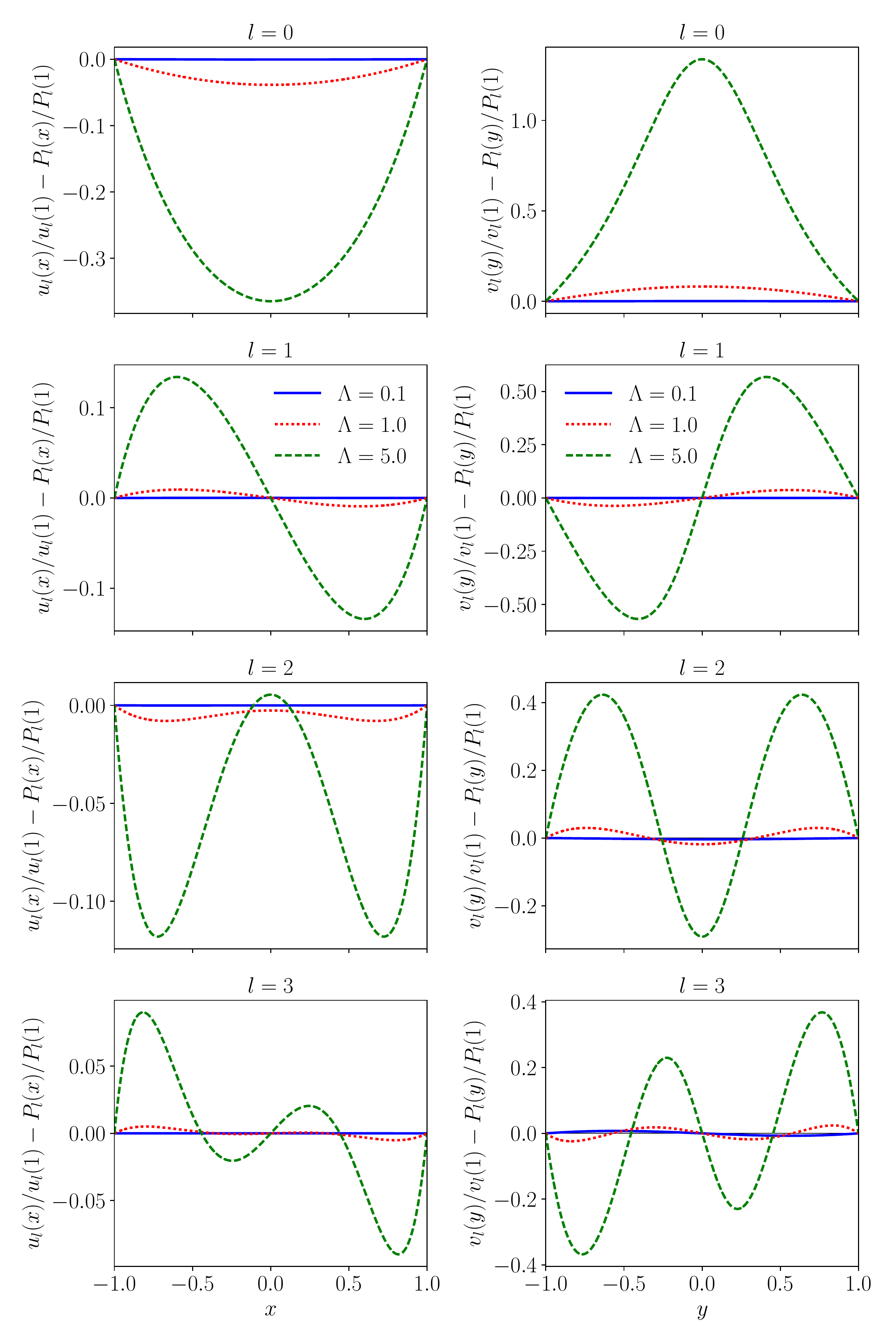}
		\end{minipage}
		\begin{minipage}{0.5\hsize}
			\centering
			\includegraphics[width=0.95\textwidth,clip]{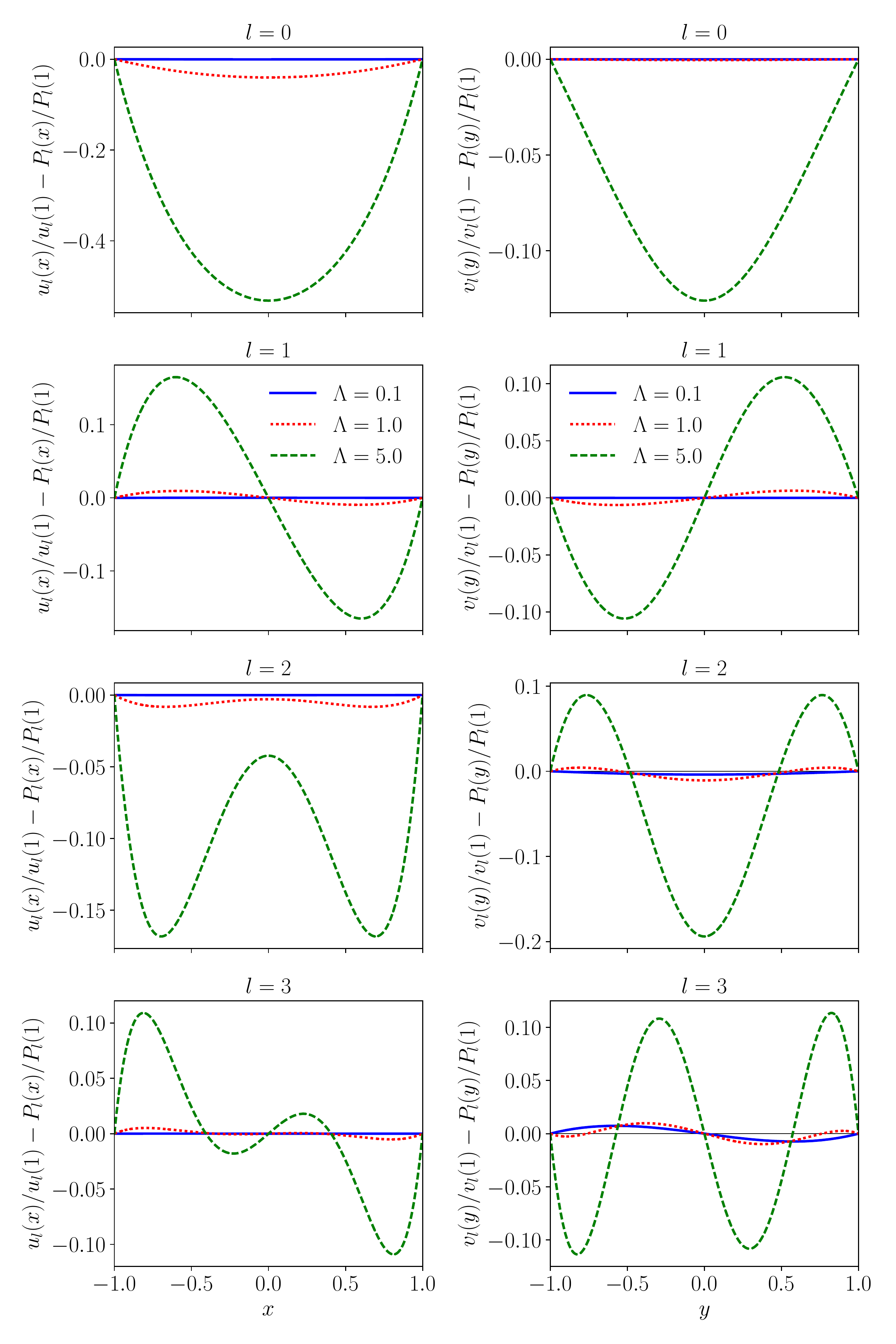}
		\end{minipage}
	\end{tabular}
	\caption{
	Basis functions of the IR for the fermionic kernel (left panel) and bosonic kernel (right panel).
	We plot the differences between the basis functions and the Legendre polynomials.
	}
	\label{fig:diff}
\end{figure*}

In Fig.~\ref{fig:diff}, we show that the basis functions $u_l(x)$ and $v_l(y)$ converge to the Legendre polynomials in the limit of $\Lambda\rightarrow 0$~\footnote{
We provide a python script for generating Fig.~\ref{fig:diff} at a public git repository~\cite{git} and in the Suppelmental Material.}.
This is clearly seen in the numerical data shown in Fig.~\ref{fig:diff}.
Due to the fast decay of the singular values $s_l$, it is numerically difficult to compute the basis functions for large $l$ accurately when $\Lambda\ll 1$.

To study the asymptotic behavior for $\Lambda\rightarrow 0$ more precisely, we algebraically expand the fermionic kernel [Eq.~(7)] in terms of the Legendre polynomials as
%in powers of $x$ and $y$ up to a very high expansion order.
%Then, we project the kernel on the 
\begin{align}
  K_\mathrm{F}(x,y) &= \sum_{l,l^\prime=0}^{N-1} K_{ll^\prime} \sqrt{\frac{2l+1}{2}} \sqrt{\frac{2l^\prime+1}{2}} P_l(x) P_l(y).~\label{eq:kernel-ll}
\end{align}
In practical, we first expand $K_\mathrm{F}(x,y)$ in powers of $x$ and $y$.
Then, we compute the expansion coefficients in Eq.~(\ref{eq:kernel-ll}) by performing the integration over $x$ and $y$.
For $N=4$, we obtained
\begin{widetext}
\begin{align}
K_{ll^\prime} &=
\left(\begin{matrix} 2 - \frac{1}{18}\Lambda^{2}+\frac{1}{300}\Lambda^{4}  & 0 & - \frac{\sqrt{5}}{45} \Lambda^{2}+\frac{\sqrt{5} }{525}\Lambda^{4} & 0\\
0 & - \frac{1}{3}\Lambda+ \frac{1}{50}\Lambda^{3}- \frac{17 }{11760}\Lambda^{5}  & 0 &\frac{\sqrt{21}}{525} \Lambda^{3} - \frac{17 \sqrt{21}}{79380} \Lambda^{5} \\
\frac{\sqrt{5}}{90}  \Lambda^{2}- \frac{\sqrt{5}}{1400}  \Lambda^{4} & 0 & \frac{1}{45}\Lambda^{2} - \frac{1}{490}\Lambda^{4}  & 0\\
0 & - \frac{\sqrt{21} }{2100}\Lambda^{3}+\frac{\sqrt{21}}{26460}\Lambda^{5}  & 0 & - \frac{1}{1050}\Lambda^{3}+ \frac{1}{8505}\Lambda^{5}\end{matrix}\right)
\end{align}
\end{widetext}
The basis functions of the IR can be computed by diagonalizing the matrix $\bK^T \bK$ ($y$ space) and the matrix $\bK \bK^T$ ($x$ space), respectively.
The singular values of $\bK$ are given by 
\begin{align}
&
\left(
\begin{matrix}
2 - \frac{1}{18}\Lambda^{2} + \frac{133 }{32400}\Lambda^{4} \\
\frac{1}{3}\Lambda - \frac{1}{50}\Lambda^{3} + \frac{4607}{2940000}\Lambda^{5} \\
\frac{1}{45}\Lambda^{2} - \frac{113}{79380} \Lambda^{4} \\
\frac{1}{1050}\Lambda^{3} - \frac{257}{4252500}\Lambda^{5} 
\end{matrix}
\right)+ \mathcal{O}\left(\Lambda^{6}\right).
\end{align}
This suggests that the $l$-th singular value is $O(\Lambda^l)$.

The matrix representation of the basis functions for $y$ reads
\begin{widetext}
\begin{align}
\left(\begin{matrix}1 + \frac{4363 }{1587600}\Lambda^{4}& 0 & \frac{\sqrt{5} }{90}\Lambda^{2} - \frac{2 \sqrt{5}}{2835} \Lambda^{4}  & 0\\
0 & 1 + \frac{14801 }{1984500} \Lambda^{4} & 0 & \frac{\sqrt{21} }{175} \Lambda^{2}- \frac{41 \sqrt{21}}{135000} \Lambda^{4} \\
%- \frac{\sqrt{5}}{5670}  \Lambda^{2}\left( 63- 4 \Lambda^{2} \right) 
- \frac{63\sqrt{5}}{5670}  \Lambda^{2}+ \frac{2\sqrt{5}}{2835}\Lambda^{4}  
& 0 & 1 - \frac{1}{3240}\Lambda^{4}  & 0\\
0 & 
%- \frac{\sqrt{21}}{945000} \Lambda^{2}\left( 5400 - 287 \Lambda^{2} \right) 
- \frac{54\sqrt{21}}{9450} \Lambda^{2} + \frac{287\sqrt{21}}{945000} \Lambda^{4}
& 0 & 1 - \frac{3}{8750}\Lambda^{4}
\end{matrix}
\right) + \mathcal{O}\left(\Lambda^{6}\right) ,
\end{align}
\end{widetext}
where the $l$-th column corresponds to the $l$-th basis function (up to sign factors).
Similarly, the basis functions for $x$ reads
\begin{widetext}
\begin{align}
\left(\begin{matrix}1 + \frac{46}{99225}\Lambda^{4} & 0 & - \frac{\sqrt{5}}{180}\Lambda^{2} + \frac{37 \sqrt{5}}{113400} \Lambda^{4} & 0\\
0 & 1 + \frac{38513}{7938000}\Lambda^{4}  & 0 & - \frac{\sqrt{21}}{700} \Lambda^{2} + \frac{97 \sqrt{21}}{2205000} \Lambda^{4} \\
%\frac{\sqrt{5} }{113400}\Lambda^{2} \left(630 - 37 \Lambda^{2} \right) 
\frac{63\sqrt{5} }{11340}\Lambda^{2} - \frac{37\sqrt{5} }{113400}\Lambda^{4} 
& 0 &  1- \frac{1}{12960} \Lambda^{4} & 0\\
0 & 
%\frac{\sqrt{21}}{2205000} \Lambda^{2} \left(3150 - 97 \Lambda^{2} \right) 
\frac{63\sqrt{21}}{44100} \Lambda^{2} - \frac{97\sqrt{21}}{2205000} \Lambda^{4}  
& 0 & 1 - \frac{3}{140000}  \Lambda^{4}
\end{matrix}\right) + \mathcal{O}\left(\Lambda^{6}\right) ,
\end{align}
\end{widetext}
One can clearly see that these basis functions converge to the Legendre polynomials as $\Lambda\rightarrow 0$. 

%\bibliography{../ref,../unpublish}

\end{document}